\UseRawInputEncoding
\documentclass[prd,noshowkeys,onecolumn,superscriptaddress]{revtex4}
\usepackage{graphicx}
\usepackage{amsmath}
\usepackage{inputenc}

\def\gap{\;\rlap{\lower 2.5pt
 \hbox{$\sim$}}\raise 1.5pt\hbox{$>$}\;}
\def\lap{\;\rlap{\lower 2.5pt
   \hbox{$\sim$}}\raise 1.5pt\hbox{$<$}\;}
\def\gsim{\;\rlap{\lower 2.5pt
 \hbox{$\sim$}}\raise 1.5pt\hbox{$>$}\;}
\def\lsim{\;\rlap{\lower 2.5pt
   \hbox{$\sim$}}\raise 1.5pt\hbox{$<$}\;}

\def\spose#1{\hbox to 0pt{#1\hss}}
\def\lta{\mathrel{\spose{\lower 3pt\hbox{$\mathchar''218$}}
     \raise 2.0pt\hbox{$\mathchar''13C$}}}
\def\gta{\mathrel{\spose{\lower 3pt\hbox{$\mathchar''218$}}
     \raise 2.0pt\hbox{$\mathchar''13E$}}}
\newcommand{\be}{\begin{equation}}
\newcommand{\ee}{\end{equation}}

\newcommand{\ls}{\mathrel{\raise1.16pt\hbox{$<$}\kern-7.0pt 
\lower3.06pt\hbox{{$\scriptstyle \sim$}}}}         
\newcommand{\gs}{\mathrel{\raise1.16pt\hbox{$>$}\kern-7.0pt 
\lower3.06pt\hbox{{$\scriptstyle \sim$}}}}         

\long\def\comment#1{}

\def\fun#1#2{\lower3.6pt\vbox{\baselineskip0pt\lineskip.9pt
  \ialign{$\mathsurround=0pt#1\hfil##\hfil$\crcr#2\crcr\sim\crcr}}}
\def\lap{\mathrel{\mathpalette\fun <}}
\def\gap{\mathrel{\mathpalette\fun >}}
\newcommand{\ba}{\begin{eqnarray}}
\newcommand{\ea}{\end{eqnarray}}

\begin{document}
\bibliographystyle{apsrev.bst}
\title{Semi-classical rotating black hole in loop quantum gravity}

\author{Zhaoyi Xu}
\email{zyxu@gzu.edu.cn}
\affiliation{College of Physics, Guizhou University, Guiyang 550025, China}
\affiliation{Key Laboratory of Particle Astrophysics,
Institute of High Energy Physics, Chinese Academy of Sciences,
Beijing 100049, China}

\begin{abstract}
In the research paper \cite{2020PhRvD.101h4001L}, the analytical solution of semi-classical rotating black holes (BH) in loop quantum gravity theory (LQG) is obtained, but an unknown function $H$ is still preserved. In this note, we obtain an expression for the unknown function $H$, which makes the space-time line element of the semi-classical rotating BH in the LQG theory sufficiently well-expressed and thus provides a basis for the study of such rotating BHs.
\end{abstract}

\maketitle{}

\section{Introduction}
\label{intro}
Quantum gravity is one of the frontiers of current physics. It attempts to reconcile the contradiction between general relativity and quantum field theory in order to achieve a consistent understanding of such questions as black holes and the origin of the universe \cite{Polchinski1,Polchinski2,Rovelli}.
Physicists have explored quantum gravity for decades and come up with loop quantum gravity theory and string theory (as well as other theories of quantum gravity)\cite{Polchinski1,Polchinski2,Rovelli}.
The main idea of loop quantum gravity theory is to quantize gravity while preserving the background dependence properties of general relativity. loop quantum gravity theory is one of the main approaches to preserve the non-perturbative properties and the background dependence \cite{2007LNP...721..185T,2003LNP...631...41T}.
From the perspective of general relativity, loop quantum gravity theory is expected to solve classical singularities problem (such as the common Big Bang naked singularities and black hole singularities).
Since general relativity is a geometric theory describing gravitational phenomena, the loop quantum gravity theory can be understood as quantizing the continuous geometric structure, making the loop quantum gravity theory also a quantum geometric theory.
​In recent decades, there has been extensive work by physicists on loop quantum gravity theory and its applications, due to the significant progress made on the problem of cosmological singularities 
\cite{2005LRR.....8...11B,2001PhRvL..86.5227B,2006PhRvD..73l4038A}.

​In recent years, black hole physics has become a testing ground for theories of quantum gravity, as scientists have made breakthroughs in the observation of binary black holes gravitational waves and shadow of supermassive black holes \cite{2016PhRvL.116f1102A,2019ApJ...875L...1E}.
​Of course, this includes testing loop quantum gravity theories using black hole gravitational waves and black hole shadow observations, which naturally makes the use of loop quantum gravity theories to study black holes an attractive problem \cite{2004PhRvD..70l4009M,2007hep.th....1239M,2006CQGra..23..391A,2006CQGra..23.5587M,2008PhRvL.101p1301G}.
Remarkably, physicists have discovered that it is possible to solve the singularity problem of spherically symmetric black holes via loop quantum gravity theory. They have found that when black holes are considered in loop quantum gravity theories, there is no singularity at the center of the black hole, while the event horizon of the black hole is preserved. This is true for both the full loop quantum gravity theory case and the semi-classical case of loop quantum gravity 
\cite{2006CQGra..23..391A}.
In the case of full-loop quantum gravity theory. In article \cite{2019arXiv191200774B}, the analytical form of the black hole metric under spherically symmetric conditions is obtained. Suddhasattwa. et.al extended it to the case of rotating black holes via the Newman-Janis method and investigated the possibility of testing loop quantum gravity in astronomical phenomena 
\cite{2021PhRvL.126r1301B}.
​In the semi-classical approximation, Leonardo Modesto reduces the action of loop quantum gravity on black holes to two effective parameters and obtains the semiclassical spherically symmetric black hole metric in loop quantum gravity theory \cite{2010IJTP...49.1649M}.
After this, Francesco Caravelli and Leonardo Modesto extended the semi-classical black hole solution to rotating black holes by the Newman-Janis method, but Azreg-Aïnou Mustapha showed that the rotating black hole metric is not a solution of Einstein's gravitational field equations \cite{2010CQGra..27x5022C,2011CQGra..28n8001A}.
In literature \cite{2020PhRvD.101h4001L}, the Newman-Janis method was used to derive the rotating black hole in the semi-classical case of loop quantum gravity theory, but they still retained an unknown function $H$, which made them not actually obtain the solution of the rotating black hole in the semi-classical case of loop quantum gravity theory.

In this work, we will obtain the analytical expression of the unknown function $H$ by approximate calculation, so that the solution of the rotating black hole in the semi-classical case of loop quantum gravity theory can be obtained.
The paper is organized as follows. In Section \ref{kerr-like}, we will construct the solution of the rotating black hole according to the Newman-Janis method, and obtain the analytic expression of the unknown function $H$. In Section \ref{proper}, we temporarily analyze and discuss the properties of rotating black holes in the semi-classical case of loop quantum gravity theory. The summary is in Section \ref{discuss}. We have used the natural system of units throughout the paper.

\section{Kerr-like semi-classical black hoe in LQG}
\label{kerr-like}
According to the research paper \cite{2010IJTP...49.1649M,2010CQGra..27x5022C,2020PhRvD.101h4001L}, the space-time line elements of spherically symmetric black holes (BH) under the semi-classical approximation in the theory of loop quantum gravity (LQG) are as follows
\begin{equation}
ds^{2}=-f(r)dt^{2}+\dfrac{1}{g(r)}dr^{2}+h(r)(d\theta^{2}+\sin^{2}\theta d\phi^{2}),
\label{metric1}
\end{equation}
where $f(r)$, $g(r)$ and $h(r)$ are the space-time metric coefficients, whose expressions are
\begin{equation}
f(r)=\frac{(r-r_{+})(r-r_{-})(r+r_{*})^{4}}{r^{4}+a^{2}_{0}},
\label{metric2}
\end{equation}
\begin{equation}
g(r)=\frac{(r-r_{+})(r-r_{-})r^{4}}{(r+r_{*})^{2}(r^{4}+a^{2}_{0})},
\label{metric3}
\end{equation}
and
\begin{equation}
h(r)=r^{2}+\frac{a^{2}_{0}}{r^{2}}.
\label{metric4}
\end{equation}
​In the metric coefficients, the constants $r_{-}$ and $r_{+}$ are the inner and outer horizons, denoted $r_{+}=2M/(1+P)^{2}$ and $r_{-}=2MP^{2}/(1+P)^{2}$, respectively. The constant in the metric coefficient $r_{*}=\sqrt{r_{+}r_{-}}$.
​The physical meaning of the parameters $a_{0}$, $M$ and $P$ can be found in \cite{2010IJTP...49.1649M}.

The energy-momentum tensor corresponding to the metric (\ref{metric1}) of a semi-classical spherically symmetric BH can be written as follows $T_{\mu\nu}=diag(-\rho,p_{r},p_{\theta},p_{\phi})$, where $\rho$ is the energy density, $p_{r}$ is the radial pressure, $p_{\theta}$ and $p_{\phi}$ is the tangential pressure. These energy density and pressure of the semi-classical BH in LQG are

\begin{equation}
-8\pi\rho=-g(r)\Big(-\frac{h^{'}g^{'}}{2hg}-\frac{1}{h}\Big)-\frac{1}{h},
\label{energy-m1}
\end{equation}

\begin{equation}
8\pi p_{r}=g(r)\Big(\frac{h^{'}f^{'}}{2hf}+\frac{1}{h}\Big)-\frac{1}{h},
\label{energy-m2}
\end{equation}

\begin{equation}
8\pi p_{\theta}=8\pi p_{\phi}=g(r)\Big[\frac{f^{''}}{2f}-\frac{(f^{'})^{2}}{4f^{2}}+\frac{g^{'}f^{'}}{4gf}+\frac{h^{'}}{4h}(\frac{f^{'}}{f}+\frac{g^{'}}{g})\Big].
\label{energy-m3}
\end{equation}

According to LQG theory, in the common phenomena of astrophysics, the quantum gravitational effect should normally be extremely tiny (which has not been observed yet), which makes the parameter values in the model considerably restricted. It is commonly believed that the parameters $P$ and $a_{0}$ of semi-classical black holes tend to zero or are much less than $1M$.
Therefore, in the LQG theory, the radial pressure and tangential pressure corresponding to the semi-classical black hole (\ref{metric1}) should be approximately equal $p_{r}=p_{\theta}=p$. In other words, the following conditions should be satisfied:
\begin{equation}
\frac{f^{''}}{2f}--\frac{(f^{'})^{2}}{4f^{2}}+\frac{1}{4gf}+\frac{h^{'}}{4h}\Big(\frac{g^{'}}{g}-\frac{f^{'}}{f}\Big)=\frac{1}{h}\Big(1-\frac{1}{g}\Big).
\label{relation1}
\end{equation}
This condition indicates that the pressure corresponding to the metric (1) of a semi-classical black hole is isotropic. Computationally, we find that the expression for the pressure $p$ is of the form:
\begin{equation}
p=\frac{(r-r_{+})(r-r_{-})(r^{4}-a^{2}_{0})r^{3}}{8\pi(r+r_{*})^{2}(r^{4}+a^{2}_{0})}\Big(\dfrac{2r-(r_{+}+r_{-})}{(r-r_{+})(r-r_{-})}+\dfrac{4}{r+r_{*}}-\dfrac{4r^{3}}{r^{4}+a^{2}_{0}}\Big)+\frac{(r-r_{+})(r-r_{-})r^{6}}{8\pi(r+r_{*})^{2}(r^{4}+a^{2}_{0})^{2}}-\dfrac{r^{2}}{8\pi(r^{4}+a^{2}_{0})}.
\label{relation2}
\end{equation}

According to the discussion in \cite{2014PhLB..730...95A}(and their references), the semi-classical spherically symmetric space-time metric (\ref{metric1}) is extended to the rotational case, and the general rotational space-time line elements are as follows
\begin{equation}
ds^{2}=-\dfrac{H}{\Sigma^{2}}\big(1-\dfrac{2\overline{f}}{\Sigma^{2}} \big)dt^{2}+\dfrac{H}{\Delta}dr^{2}-\dfrac{4a\overline{f}\sin^{2}\theta H}{\Sigma^{4}}dtd\phi+H d\theta^{2}+\dfrac{H A \sin^{2}\theta}{\Sigma^{4}}d\phi^{2},
\label{NJA10}
\end{equation}
where $k(r)=h(r)\sqrt{f(r)/g(r)}$, $2\overline{f}=k(r)-h(r)f(r)$, $\Delta(r)=h(r)f(r)+a^{2}$, $A=(k(r)+a^{2})^{2}-a^{2}\Delta \sin^{2}\theta$, $\Sigma^{2}=k(r)+a^{2}\cos^{2}\theta$ and $H=H(r,\theta,a)$ is unknown function. 
In order to find the specific expression of the unknown function $H$, we need to use the rotational condition $G_{r\theta}=0$ and the Einstein field equation $G_{\mu\nu}=8\pi T_{\mu\nu}$. By substituting the space-time line element (\ref{NJA10}) into these two conditions, we can obtain the following system of equations
\begin{equation}
(k+a^{2}y^{2})^{2}(3H_{,r}H_{,y^{2}}-2HH_{,ry^{2}})=3a^{2}k_{,r}H^{2},
\label{NJA11}
\end{equation}
\begin{equation}
H[k^{2}_{,r}+k(2-k_{,rr})-a^{2}y^{2}(2+k_{,rr})]+(k+a^{2}y^{2})(4y^{2}H_{,y^{2}}-k_{,r}H_{,r})=0,
\label{NJA12}
\end{equation}
where $H_{,ry^{2}}=\partial^{2}H/\partial r \partial y^{2}$, $k_{,r}=\partial k(r)/\partial r$ and $y=\cos\theta$. 
In Reference \cite{2020PhRvD.101h4001L}, authors discuss the properties of rotating BHs while retaining the unknown function $H$. In fact, we can figure out the expression for the unknown function $H$. For the spherically symmetric metric of $f(r)\neq g(r)$, the solution of the unknown function $H$ can be found under the condition $P_{r}\simeq P_{\theta}\simeq P_{\phi}=p$. In the literature \cite{2014PhLB..730...95A}, the author gives the strict solution of the equations (\ref{NJA11}) and (\ref{NJA12}), whose analytical form is as follows
\begin{equation}
H=r^{2}+p^{2}+a^{2}y^{2}=r^{2}+p^{2}+a^{2}\cos^{2}\theta,
\label{NJA15}
\end{equation}
​here $p$ is the pressure and is given by the expression (\ref{relation2}).
To sum up, the analytical expressions of the unknown functions and parameters in the rotating BH (\ref{NJA10}) can be obtained by proper arrangement. 
\begin{equation}
H=\Big[\frac{(r-r_{+})(r-r_{-})(r^{4}-a^{2}_{0})r^{3}}{8\pi(r+r_{*})^{2}(r^{4}+a^{2}_{0})}\Big(\dfrac{2r-(r_{+}+r_{-})}{(r-r_{+})(r-r_{-})}+\dfrac{4}{r+r_{*}}-\dfrac{4r^{3}}{r^{4}+a^{2}_{0}}\Big)+\frac{(r-r_{+})(r-r_{-})r^{6}}{8\pi(r+r_{*})^{2}(r^{4}+a^{2}_{0})^{2}}-\dfrac{r^{2}}{8\pi(r^{4}+a^{2}_{0})}\Big]^{2}$$$$
+r^{2}+a^{2}\cos^{2}\theta,
\label{NFW-BH1}
\end{equation}
\begin{equation}
k(r)=\Big(r^{2}+\frac{a^{2}_{0}}{r^{2}}\Big)\Big(1+\frac{r_{*}}{r}\Big)^{2}(r+r_{*}),~~~~~\Sigma^{2}=\Big(r^{2}+\frac{a^{2}_{0}}{r^{2}}\Big)\Big(1+\frac{r_{*}}{r}\Big)^{2}(r+r_{*})+a^{2}\cos^{2}\theta,
\label{NFW-BH2}
\end{equation}
\begin{equation}
2\overline{f}=\Big(r^{2}+\frac{a^{2}_{0}}{r^{2}}\Big)\Big(1+\frac{r_{*}}{r}\Big)^{2}(r+r_{*})-(1-\frac{r_{+}}{r} )(1-\frac{r_{-}}{r})(r+r_{*})^{4},~~~~~~
\Delta(r)=(1-\frac{r_{+}}{r} )(1-\frac{r_{-}}{r})(r+r_{*})^{4}+a^{2},
\label{NFW-BH3}
\end{equation}
\begin{equation}
A=\Big(\Big(r^{2}+\frac{a^{2}_{0}}{r^{2}}\Big)\Big(1+\frac{r_{*}}{r}\Big)^{2}(r+r_{*})+a^{2}\Big)^{2}-a^{2}\sin^{2}\theta\Big((1-\frac{r_{+}}{r} )(1-\frac{r_{-}}{r})(r+r_{*})^{4}+a^{2}\Big),
\label{NFW-BH4}
\end{equation}

\section{Some properties and discussion}
\label{proper}
For the semi-classical BH metric (\ref{NJA10}) in LQG theory, we would temporarily discuss and analyze its basic properties.
Firstly, the structure of the BH event horizon determined by the semi-classical BH metric (\ref{NJA10}) is analyzed.
According to the BH event horizon structure equation $g^{rr}=-\frac{H}{\Delta}=0$, it can be found that the event horizon structure of the BH is only determined by the expression of $\Delta$, and has no relation with the expression of function $H$.
Therefore, the expression of function $H$ would not modify the analysis of BH event horizon structure in reference \cite{2020PhRvD.101h4001L}. Details of the event horizon structure and properties of BH (\ref{NJA10}) have been thoroughly discussed in reference \cite{2020PhRvD.101h4001L}.
Secondly, for the properties of the stationary limit surfaces and ergosphere of the semi-classical BH (\ref{NJA10}), since the inner and outer stationary limit surfaces are determined by the equation $g^{tt}=-\dfrac{H}{\Sigma^{2}}\big(1-\dfrac{2\overline{f}}{\Sigma^{2}}\big)=0$, the structure of the function $H$ does not modify the properties of the inner and outer stationary limit surfaces. A detailed analysis of the inner and outer stationary limit surfaces can be found in \cite{2020PhRvD.101h4001L}.
The boundary of the ergosphere is determined by the outer horizon of the BH and the outer stationary limit surface, so the size of the ergosphere region has no relationship with the function $H$, and its analysis is still exactly the same as that in reference \cite{2020PhRvD.101h4001L}. 
Finally, the singularity problem of semi-classical BH metric (\ref{NJA10}). The singularity of the BH can be analyzed by computing the Kretschmann scalar $R$, which is
\begin{equation}
R=\dfrac{2}{\Big[\Big(r^{2}+\frac{a^{2}_{0}}{r^{2}}\Big)\Big(1+\frac{r_{*}}{r}\Big)^{2}(r+r_{*})+a^{2}\cos^{2}\theta \Big]^{3}}\times\Big[r^{2}+\Big[\frac{(r-r_{+})(r-r_{-})(r^{4}-a^{2}_{0})r^{3}}{8\pi(r+r_{*})^{2}(r^{4}+a^{2}_{0})}\Big(\dfrac{2r-(r_{+}+r_{-})}{(r-r_{+})(r-r_{-})}+\dfrac{4}{r+r_{*}}$$$$
-\dfrac{4r^{3}}{r^{4}+a^{2}_{0}}\Big)+\frac{(r-r_{+})(r-r_{-})r^{6}}{8\pi(r+r_{*})^{2}(r^{4}+a^{2}_{0})^{2}}-\dfrac{r^{2}}{8\pi(r^{4}+a^{2}_{0})}\Big]^{2}+a^{2}(2-\cos^{2}\theta)-\frac{2(r-r_{+})(r-r_{-})(r+r_{*})^{4}}{r^{4}+a^{2}_{0}}\Big]\times  $$$$\Big[\frac{(r-r_{+})(r-r_{-})(r^{4}-a^{2}_{0})r^{3}}{8\pi(r+r_{*})^{2}(r^{4}+a^{2}_{0})}\Big(\dfrac{2r-(r_{+}+r_{-})}{(r-r_{+})(r-r_{-})}+\dfrac{4}{r+r_{*}}-\dfrac{4r^{3}}{r^{4}+a^{2}_{0}}\Big)+\frac{(r-r_{+})(r-r_{-})r^{6}}{8\pi(r+r_{*})^{2}(r^{4}+a^{2}_{0})^{2}}-\dfrac{r^{2}}{8\pi(r^{4}+a^{2}_{0})}\Big]^{2}$$$$
-\dfrac{2}{\Big(r^{2}+\frac{a^{2}_{0}}{r^{2}}\Big)\Big(1+\frac{r_{*}}{r}\Big)^{2}(r+r_{*})+a^{2}\cos^{2}\theta}\times\Big[ \frac{2(r+r_{*})^{4}+8(2r-r_{+}-r_{-})(r+r_{*})^{3}+12(r-r_{+})(r-r_{-})(r+r_{*})^{2}}{r^{4}+a^{2}_{0}}+$$$$
\frac{32(r-r_{+})(r-r_{-})(r+r_{*})^{4}r^{6}}{(r^{4}+a^{2}_{0})^{3}}
-\frac{12(r-r_{+})(r-r_{-})(r+r_{*})^{4}r^{3}+8(2r-r_{+}-r_{-})(r+r_{*})^{4}r^{3}+32(r-r_{+})(r-r_{-})(r+r_{*})^{3}r^{3}}{(r^{4}+a^{2}_{0})^{2}} \Big].
\label{Kretschmann scalar}
\end{equation}
​It follows from this expression that $R$ is finite in all spacetime regions. 
The specific properties of the function $H$ do not affect the qualitative discussion of the singularity of the BH, but can modify the magnitude of the value of $R$, and thus alter the related physical properties.

For the semi-classical BH metric (\ref{NJA10}), although the previous discussion is all about the space-time metric in vacuum, it can actually be equivalent to the curvature of space-time caused by some material fluid, which allows us to understand the physical properties of the BH by calculating the energy-momentum tensor component of the metric (\ref{NJA10}). By a series of calculations, we find that its non-zero component are
\begin{equation}
\rho(Kerr-like BH)=\frac{1}{\Sigma^{6}}\Big[-6f+r^{2}+p^{2}+a^{2}(2-\cos^{2}\theta)\Big]p^{2}-\frac{2}{\Sigma^{4}}\Big(rf^{'} -f\Big),
\label{EMT1}
\end{equation}
\begin{equation}
p_{r}(Kerr-like BH)=\frac{1}{\Sigma^{6}}\Big[2f+r^{2}+p^{2}+a^{2}\cos^{2}\theta\Big]p^{2}-\frac{2}{\Sigma^{4}}\Big(rf^{'} -f\Big),
\label{EMT2}
\end{equation}
\begin{equation}
p_{\theta}(Kerr-like BH)=\frac{2(r^{2}+a^{2}\cos^{2}\theta)}{\Sigma^{6}}f-\frac{1}{\Sigma^{4}}\Big[2rf^{'}+p^{2}\Big]+\frac{1}{\Sigma^{2}}f^{''} ,
\label{EMT3}
\end{equation}
\begin{equation}
p_{\phi}(Kerr-like BH)=\frac{2}{\Sigma^{6}}[(r^{2}+a^{2}\cos^{2}\theta)f-2a^{2}\sin^{2}\theta p^{2}]--\frac{1}{\Sigma^{4}}\Big[2rf^{'}+p^{2}\Big]+\frac{1}{\Sigma^{2}}f^{''},
\label{EMT4}
\end{equation}
where $f^{'}$ and $f^{''}$ are
\begin{equation}
f^{'}=\frac{(2r-r_{+}-r_{-})(r+r_{*})^{4}}{r^{4}+a^{2}_{0}}+\frac{4(r-r_{+})(r-r_{-})(r+r_{*})^{3}}{r^{4}+a^{2}_{0}}-\frac{4(r-r_{+})(r-r_{-})(r+r_{*})^{4}r^{3}}{(r^{4}+a^{2}_{0})^{2}},
\label{f1}
\end{equation}
\begin{equation}
f^{''}=\frac{2(r+r_{*})^{4}+8(2r-r_{+}-r_{-})(r+r_{*})^{3}+12(r-r_{+})(r-r_{-})(r+r_{*})^{2}}{r^{4}+a^{2}_{0}}+\frac{32(r-r_{+})(r-r_{-})(r+r_{*})^{4}r^{6}}{(r^{4}+a^{2}_{0})^{3}}$$$$
-\frac{12(r-r_{+})(r-r_{-})(r+r_{*})^{4}r^{3}+8(2r-r_{+}-r_{-})(r+r_{*})^{4}r^{3}+32(r-r_{+})(r-r_{-})(r+r_{*})^{3}r^{3}}{(r^{4}+a^{2}_{0})^{2}},
\label{f2}
\end{equation}

Through these expressions of energy density and pressure, it can be understood that in the semi-classical rotating BH space-time of LQG theory, the equivalent fluids corresponding to BH (\ref{NJA10}) do not satisfy the isotropic conditions. 
From the previous discussion, we know that the semi-classical BH metric (1) is approximately isotropic in the case of spherically symmetric case. Since the BH metric (\ref{NJA10}) takes into account the BH spin, the isotropic condition is destroyed, which makes the physical properties of BH (\ref{NJA10}) more complicated.

\section{Summary}
\label{discuss}
In the research paper \cite{2020PhRvD.101h4001L}, the authors used the Newman-Janis method to derive the semi-classical rotating BH metric in the theory of LQG, but there is still an unknown function $H$ in their semi-classical rotating BH metric. In this comment we give the analytical expression of this unknown function $H$, which makes it possible to write the mathematical form of the metric of semi-classical rotating BHs in the theory of LQG  altogether.

The conclusions in the research paper \cite{2020PhRvD.101h4001L} are generally unaffected, and our work is only to improve the space-time metric so that some of the conclusions can be changed from qualitative to quantitative calculations.

\begin{acknowledgments}
We acknowledge the anonymous referee for a constructive report that has significantly improved this paper. We acknowledge the  Special Natural Science Fund of Guizhou University (grant
No. X2020068) and the financial support from the China Postdoctoral Science Foundation funded project under grants No. 2019M650846.
\end{acknowledgments}

\end{document}